\documentclass[runningheads]{llncs}
\usepackage{cite}
\usepackage{amsmath,amssymb,amsfonts}
\usepackage{algorithmic}
\usepackage{graphicx}
\usepackage{textcomp}
\usepackage{comment}
\usepackage{times}
\usepackage{bmpsize}
\usepackage{xcolor}
\usepackage{lipsum}
\usepackage{todonotes}
\usepackage{wrapfig}
\usepackage{flushend}
\usepackage{multirow}
\usepackage[title]{appendix}

\usepackage[colorlinks=true,urlcolor=black]{hyperref}
\def\BibTeX{{\rm B\kern-.05em{\sc i\kern-.025em b}\kern-.08em
    T\kern-.1667em\lower.7ex\hbox{E}\kern-.125emX}}
    
\usepackage{booktabs}   
\usepackage{subcaption} 

\usepackage[ruled,vlined,linesnumbered]{algorithm2e}

\begin{document}

\title{Verifying Neural Networks Against Backdoor Attacks}

\author{Long H. Pham and Jun Sun}
\institute{Singapore Management University}


\maketitle

\begin{abstract}
Neural networks have achieved state-of-the-art performance in solving many problems, including many applications in safety/security-critical systems. 
Researchers also discovered multiple security issues associated with neural networks. One of them is backdoor attacks, i.e., a neural network may be embedded with a backdoor such that a target output is almost always generated in the presence of a trigger. Existing defense approaches mostly focus on detecting whether a neural network is `backdoored' based on heuristics, e.g., activation patterns. To the best of our knowledge, the only line of work which certifies the absence of backdoor is based on randomized smoothing, which is known to significantly reduce neural network performance. In this work, we propose an approach to verify whether a given neural network is free of backdoor with a certain level of success rate. Our approach integrates statistical sampling as well as abstract interpretation.
The experiment results show that our approach effectively verifies the absence of backdoor or generates backdoor triggers. 
\end{abstract}

\section{Introduction}
\label{intro}

Neural networks
gradually become an essential component in many real-life systems, e.g., face recognition~\cite{parkhi2015deep}, medical diagnosis~\cite{li2014medical}, as well as auto-driving car~\cite{bojarski2016end}. Many of these systems are safety and security-critical. 
In other words, it is expected that the neural networks used in these systems should not only operate correctly but also satisfy security requirements, i.e., they must sustain attacks from malicious adversaries.

Researchers have identified multiple ways of attacking neural networks, including adversarial attacks~\cite{szegedy2013intriguing}, backdoor attacks~\cite{gu2017badnets}, and so on. Adversarial attacks apply a small perturbation (e.g., modifying few pixels in an image input) to a given input (which is often unrecognizable under human inspection) and cause the neural network to generate a wrong output. To mitigate adversarial attacks, many approaches have been proposed, including robust training~\cite{mirman2018differentiable,dong2018boosting}, run-time adversarial sample detection~\cite{wang2019adversarial}, and robustness certification~\cite{gehr2018ai2}. The most relevant to this work is robustness certification, which aims to verify that a neural network satisfies local robustness, i.e., perturbation within a region (e.g., an $L_\infty$ norm) around an input does not change the output. The problem of local robustness certification has been extensively studied in recent years and many methods and tools have been developed~\cite{katz2017reluplex,huang2017safety,wang2018efficient,gehr2018ai2,wang2018formal,singh2018fast,singh2019abstract,singh2018boosting,singh2019beyond}.  

Backdoor attacks work by embedding a `backdoor' in the neural network so that the neural network works as expected with normal inputs and outputs a specific target output in the presence of a backdoor trigger. For instance, given a `backdoored' image classification network, any image which contains the backdoor trigger will be (highly likely) assigned a specific \emph{target label} chosen by the adversary, regardless of the content of the image.
The backdoor trigger can be embedded either through poisoning the training set~\cite{gu2017badnets} or modifying a trained neural network directly~\cite{liu2017trojaning}. 
It is easy to see that backdoor attacks raise serious security concerns. For instance, the adversaries may use a trigger-containing (a.k.a.~`stamped') image to fool a face recognition system and pretend to be someone with high authority~\cite{chen2017targeted}. Similarly, a stamped image may be used to trick an auto-driving system to misidentify street signs and act hazardously~\cite{gu2017badnets}. 


There are multiple active lines of research related to backdoor attacks, e.g., on different ways of conducting backdoor attacks~\cite{gu2017badnets,liu2020reflection}, different ways of detecting the existence of backdoor~\cite{liu2017trojaning,gao2019strip,wang2019neural,chen2018detecting,liu2019abs} or mitigating backdoor attacks~\cite{liu2018fine}. 
Existing approaches are however not capable of certifying the absence of backdoor. 
To the best of our knowledge, the only work that is capable of certifying the absence of backdoor is the work reported in~\cite{wang2020certifying} which is based on the randomized smoothing during training. Their approach has a huge cost in terms of model accuracy and even the authors (in the abstract of~\cite{wang2020certifying}) are calling for alternative approaches for ``certifying robustness against backdoor attacks''. 

In this work, we propose a method to verify the absence of backdoor attack  with a certain level of success rate (since backdoor attacks in practice are rarely perfect~\cite{gu2017badnets,liu2020reflection}). Given a neural network and a constraint on the backdoor trigger (e.g., its size), our method is a combination of statistical sampling and deterministic neural network verification techniques (based on abstract interpretation). If we fail to verify the absence of backdoor (due to over-approximation), an optimization-based method is developed to generate concrete backdoor triggers. 

We conduct experiments on multiple neural networks trained to classify images in the MNIST dataset. 
These networks are trained with different types of activation functions, including ReLU, Sigmoid, and Tanh. We verify the absence of backdoor with different settings. The experiment results show that we can verify most of the benign neural networks. Furthermore, we can successfully generate backdoor triggers for neural networks trained with backdoor attack. 
A slightly surprising result is that we successfully generate backdoor triggers for some of the supposedly benign networks with a reasonably high success rate.

The remaining of the paper is organized as follows. In Section~\ref{overview}, we define our problem. 
In Section~\ref{sec:approach}, we present the details of our approach. 
We show the experiment results in Section~\ref{eval}. Section~\ref{related} reviews related work and finally, Section~\ref{conclusion} concludes.
\section{Problem Definition}
\label{overview}


In the following, our discussion focuses on the image domain, in particular, on image classification neural networks. It should be noted that our approach is not limited to the image domain.  
In general, an image can be represented as a three-dimensional array with shape $(c,h,w)$ where $c$ is the number of channels (i.e., $1$ for grayscale images and $3$ for color images); $h$ is the height (i.e., the number of rows); and $w$ is the width (i.e., the number of columns) of the image. Each element in the array is a byte value (i.e., from $0$ to $255$) representing a feature of the image. When an image is used in a classification task with a neural network, its feature values are typically normalized into floating-point numbers (e.g., dividing the original values by $255$ to get normalized values from $0$ to $1$). Moreover, the image is transformed into a vector with size $m = c \times h \times w$. In this work, we use the three-dimensional form and the vector form of an image interchangeably. The specific form which we use should be clear from the context.

\begin{figure*}[t]
\centering
\includegraphics[scale=0.08]{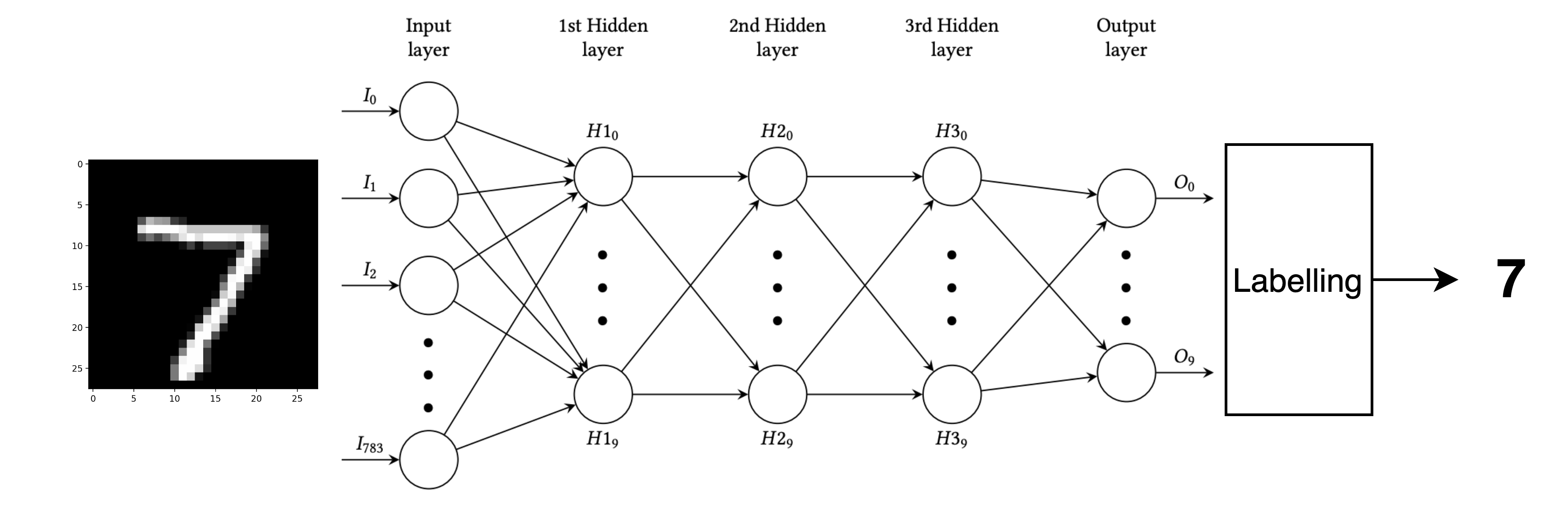}
\caption{An example of image classification with neural network}
\label{fig:network}
\end{figure*}

Given a tuple $(c_i, h_i, w_i)$ representing an index in the three-dimensional form, it is easy to compute the according index $i$ in the vector form using the formula: $i = c_i \times h \times w + h_i \times w + w_i$. Similarly, given an index $i$ in the vector form, we compute the tuple $(c_i, w_i, h_i)$ representing the index in the three-dimensional form as follows.
\begin{align*}
    c_i & = i \div (h \times w) \\
    h_i & = (i - c_i \times h \times w) \div w \\
    w_i & = i - c_i \times h \times w - h_i \times w
\end{align*}
An image classification task is to label a given image with one of the pre-defined labels automatically. Such tasks are often solved using neural networks. Fig.~\ref{fig:network} shows the typical workflow of an image classification neural network.
The task is to assign a label (i.e., from 0 to 9) to a handwritten digit image. Each input is a grey-scale image with $1 \times 28 \times 28 = 784$ features.

In this work, we focus on fully connected neural networks and convolutional neural networks, which are composed of multiple layers of neurons.
The layers include an input layer, a set of hidden layers, and an output layer.
The number of neurons in the input layer equals the number of features in the input image.
The number of neurons in the output layer equals the number of labels in the classification problem.
The number of hidden layers as well as the number of neurons in these layers are flexible.
For instance, the network in Fig.~\ref{fig:network} has three hidden layers, each of which contains 10 neurons.

The input layer simply applies an identity transformation on the vector of the input image.
Each hidden layer transforms its input vector (i.e., the output vector of the previous layer) and produces an output vector for the next layer.
Each hidden layer applies two different types of transformations, i.e., the first is an affine transformation and the second is an activation function transformation. Formally, the two transformations of a hidden layer can be defined as: 
$\vec{y} = \sigma(A * \vec{x} + B)$ where $\vec{x}$ is the input vector, $A$ is the weight matrix, $B$ is the bias vector of the affine transformation, $*$ is the matrix multiplication, $\sigma$ is the activation function, and $\vec{y}$ is the output vector of the layer.
The most popular activation functions include ReLU, Sigmoid, and Tanh.
The output layer applies a final affine transformation to its input vector and produces the output vector of the network.
A labelling function $L(\vec{y}) = {\arg\max}_i~\vec{y}$ is then applied on the output vector to return the index of the label with the highest value in $\vec{y}$. 

The weights and biases used in the affine transformations are considered parameters of the neural network.
In this work, we focus on pre-trained networks, i.e., the weights and biases of the networks are already fixed. Formally,
a neural network is a function $N : R^{m} \rightarrow R^{n} = f_k \circ \cdots f_i \cdots \circ f_0$ where $m$ is the number of features in the input; $n$ is the number of labels;
each component $f_i$ where $0 < i < k$ is a function as shown in (2) (i.e., the composition of the affine function and the activation function of the $i$-th hidden layer); $f_0$ is the identity transformation of the input layer; and $f_k$ is the last affine transformation of the output layer. \\

\begin{figure}[t]
\centering
\begin{subfigure}{0.48\textwidth}
  \centering
  \includegraphics[width=\linewidth]{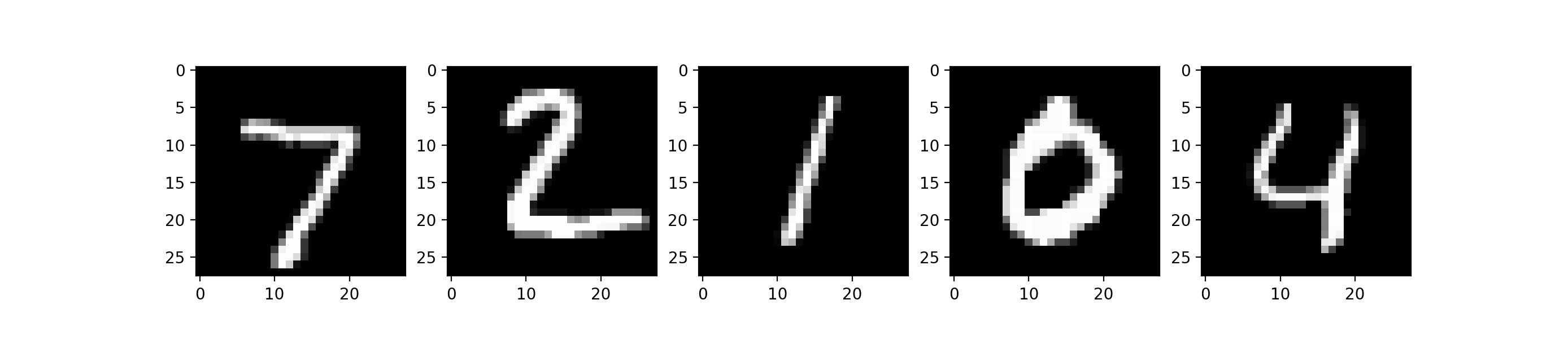}
  \caption{Original}
  \label{fig:o_imgs}
\end{subfigure}
\begin{subfigure}{0.48\textwidth}
  \centering
  \includegraphics[width=\linewidth]{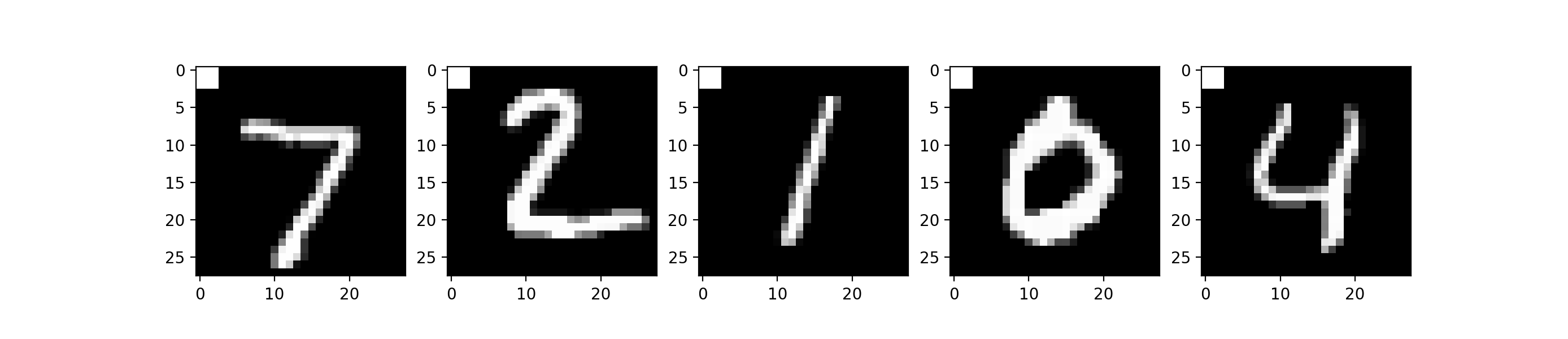}
  \caption{Stamped}
  \label{fig:s_imgs}
\end{subfigure}
\caption{Some examples of original images and stamped images}
\label{fig:imgs}
\end{figure}

\noindent \emph{Backdoor Attacks}
In~\cite{gu2017badnets}, Gu~\emph{et al.} show that neural networks are subject to backdoor attacks. Intuitively, the idea is that an adversary may introduce a backdoor into the network, for instance, by poisoning the training set. To do that, the adversary starts with choosing a pattern, i.e., a backdoor trigger, and stamps the trigger on a set of samples in the training set (e.g., 20\%). Fig.~\ref{fig:s_imgs} shows some stamped images, which are obtained by stamping a trigger to the original images in Fig.~\ref{fig:o_imgs}. Note that the trigger is a small white square at the top-left corner of the image.
A pre-defined target label is the ground truth label for the stamped images.
The poisoned training set is then used to train the neural network. The result is a backdoored network that performs normally on clean images (i.e., images without the trigger) but likely assigns the target label to any image which is stamped with the trigger. Besides poisoning the training set, a backdoor can also be introduced by modifying the parameters of a trained neural network directly~\cite{liu2017trojaning}. 

\begin{definition}[Backdoor trigger] 
Given a neural network for classifying images with shape $(c, h, w)$, a backdoor trigger is any image $S$ with shape $(c_s, h_s, w_s)$ such that $c_s = c$, $h_s \leq h$, and $w_s \leq w$.
\end{definition}
Formally, a backdoor trigger is any stamp that has the same number of channels. 
Obviously, replacing an input image entirely with a backdoor image with the same size is hardly interesting in practice. Thus, we often limit the size of the trigger in practice.  
Note that the trigger can be stamped anywhere on the image.
In this work, we assume the same trigger is used to attack all images, i.e., the same stamp is stamped at the same position given any input. In other words, we do not consider input-specific triggers, i.e., the triggers that are different for different images. While some forms of input-specific triggers (e.g., adding a specific image filter or stamping the trigger at selective positions of a given image~\cite{liu2020reflection,chen2017targeted}) can be supported by modeling the trigger as a function of the original image, we do not regard general input-specific triggers to be within the scope of this work. Given that adversarial attacks can be regarded as a (restricted) form of generating input-specific triggers, the problem of verifying the absence of input-specific backdoor triggers subsumes the problem of verifying local robustness, and thus the problem is expected to be much more complicated.

Given a trigger with shape $(c_s, h_s, w_s)$, let $(h_p, w_p)$ be the position of the top-left corner of the trigger s.t.~$h_p + h_s \leq h$ and $w_p + w_s \leq w$. Given an image $I$ with shape $(c, h, w)$, a backdoor trigger $S$ with shape $(c_s, h_s, w_s)$, and a trigger position $(h_p, w_p)$, a stamped image, denoted as $I_s$, is defined as follows.
\begin{align*}
I_s[c_i, h_i, w_i] = \left\{
\begin{array}{l}
S[c_i, h_i - h_p, w_i - w_p] ~\mbox{if $h_p \leq h_i < h_p + h_s \land w_p \leq w_i < w_p + w_s$} \\
I[c_i, w_i, h_i] \quad\quad\quad\quad\quad\quad~~ \mbox{otherwise}
\end{array}
\right.
\end{align*}
Intuitively, in the stamped image, the pixels of the stamp replace those corresponding pixels in the original image. 

Given a backdoored network, an adversary can perform an attack by feeding an image stamped with the backdoor trigger to the network and expecting the network to classify the stamped image with the target label. Ideally, given any stamped image, an attack on a backdoored network should result in the target label. In practice, experiment results from existing backdoor attacks~\cite{gu2017badnets,chen2017targeted,liu2020reflection} show that this is not always the case, i.e., some stamped images may not be classified with the target label. Thus, given a neural network $N$, a backdoor trigger $S$, a target label $t_s$, we say that $S$ has a success rate of $\theta$ if and only if there exists a position $(h_p, w_p)$ such that the probability of having $L(N(I_s)) = t_s$ for any $I$ in a chosen test set is $\theta$.

We are now ready to define the problem. 
\emph{Given a neural network $N$, a probability of $\theta$ and a trigger shape $(c_s, h_s, w_s)$, the problem of verifying the absence of a backdoor attack with a success rate of $\theta$ against $N$ is to show that there does not exist a backdoor attack on $N$ which has a success rate of at least $\theta$}. 

\section{Verifying Backdoor Absence}
\label{sec:approach}

In this section, we present our approach for addressing the problem. 
\subsection{Overall Algorithm}
The overall approach is shown in Algorithm~\ref{alg:overall}. The inputs include the network $N$, the required success rate $\theta$, a parameter $K$ representing the sampling size, the trigger shape $(c_s, h_s, w_s)$, the target label $t_s$, as well as multiple parameters for hypothesis testing (i.e., a type I error $\alpha$, a type II error $\beta$, and a half-width of the indifference region $\delta$). The idea is to apply hypothesis testing, i.e., the SPRT algorithm~\cite{agha2018survey}, with the following two mutually exclusive hypotheses. 
\begin{itemize}
    \item $H_0$: The probability of not having an attack on a set of $K$ randomly selected images is more than $1-\theta^K$.
    \item $H_1$: The probability of not having an attack on a set of $K$ randomly selected images is no more than $1-\theta^K$.
\end{itemize}

In the algorithm, variable $n$ and $z$ record the number of times a set of $K$ random images is sampled and is shown to be free of a backdoor with a 100\% success rate respectively. Note that function $VerifyX$ returns SAFE only if there is no backdoor attack on a set of given images $X$ with 100\% success rate, i.e., $L(N(I_s)) = t_s$ for all $I \in X$. It may also return a concrete trigger which successfully attacks every image in $X$. The details of algorithm $VerifyX$ is presented in Section~\ref{sec:verifyX}.

The loop from lines 4 to 15 in Algorithm~\ref{alg:overall} keeps randomly selecting and verifying a set of $K$ images using algorithm $verifyX$ until one of the two hypotheses is accepted according to the criteria set by the parameters $\alpha$ and $\beta$ based on the SPRT algorithm. Furthermore, whenever a trigger is returned by algorithm $verifyX$ at line 9, we check whether the trigger reaches the required success rate on the test set, and return UNSAFE if it does. Note that when $H_0$ is accepted, we return SAFE, i.e., we successfully verify the absence of a backdoor attack with a success rate of at least $\theta$. 
When $H_1$ is accepted, we return UNKNOWN.  

\begin{algorithm}[t]
\caption{$verifyPr(N, \theta, K, (c_s, h_s, w_s), t_s, \alpha, \beta, \delta)$}
\label{alg:overall}
let $n \leftarrow 0$ be the number of times $verifyX$ is called\;
let $z \leftarrow 0$ be the number of times $verifyX$ returns SAFE\;
let $p_0 \leftarrow (1 - \theta^K) + \delta$, $p_1 \leftarrow (1 - \theta^k) - \delta$\;
\While {true} {
  $n \leftarrow n+1$\;
  randomly select a set of images $X$ with size $K$\;
  \If {$verifyX(N, X, (c_s, h_s, w_s), t_s)$ returns SAFE}{
    $z \leftarrow z+1$;
  }
  \ElseIf {$verifyX(N, X, (c_s, h_s, w_s), t_s)$ returns UNSAFE}{
    \If{the generated trigger satisfies the success rate}{
        \Return UNSAFE;
    }
  }
    \If {$\frac{p_1^z}{p_0^z} \times \frac{(1 - p_1)^{n-z}}{(1 - p_0)^{n - z}} \le \frac{\beta}{1-\alpha}$}{
      \Return SAFE; // Accept $H_0$
    }
    \ElseIf {$\frac{p_1^z}{p_0^z} \times \frac{(1 - p_1)^{n-z}}{(1 - p_0)^{n - z}} \ge \frac{1-\beta}{\alpha}$}{
      \Return UNKNOWN; // Accept $H_1$
    }
}
\end{algorithm}

Apart from the success rate $\theta$ and parameters for hypothesis testing, Algorithm~\ref{alg:overall} has a particularly interesting parameter $K$, i.e., the number of images to draw at random each time. On the one hand, if $K$ is set to be small, such as 1, it is very likely algorithm $verifyX$ invoked at line 9 will return UNSAFE since it is often possible to attack a small set of images as demonstrated by many adversarial attack methods~\cite{goodfellow2014explaining,papernot2016limitations,carlini2017towards}, i.e., changing a few pixels of an image changes the output of a neural network. As a result, hypothesis $H_1$ is accepted and nothing can be concluded. On the other hand, if $K$ is set to be large, such as 10000, due to the complexity of algorithm $verifyX$ (see Section~\ref{sec:verifyX}), it is likely that it will timeout and thus return UNKNOWN, which leads to inclusion as well. Furthermore, when $K$ is large, $1-\theta^K$ will be close to 1 and, as a result, many rounds are needed to accept $H_0$ even if algorithm $verifyX$ returns SAFE. It is thus important to find an effective $K$ value to balance the two aspects. We identify the value of $K$ empirically in Section~\ref{eval} and aim to study the problem in the future.

Take as an example the network shown in Fig.~\ref{fig:network} which is a feed-forward neural network built with the ReLU activation function and three hidden layers. 
We aim to verify the absence of a backdoor attack with a success rate of $0.9$. We take 10000 images of the MNIST test set to evaluate the success rate of a trigger. We set the parameters in Algorithm~\ref{alg:overall} as follows: $K = 5$ and $\alpha = \beta = \delta = 0.01$. For the target label 0, after 95 rounds, we have enough evidence to accept the hypothesis $H_0$, which means we have evidence that there is no backdoor attack on the network with the target label 0 and a success rate of at least $0.9$. We have similar results for other target labels, although more rounds of tests are required for labels 2, 3, 5, and 8 (i.e., 98 rounds for label 8, 100 rounds for label 3, 117 rounds for label 5, and 188 rounds for label 2).

\subsection{Verifying Backdoor Absence against a Set of Images}
\label{sec:verifyX}
Next, we present the details of algorithm $verifyX$. The inputs include the neural network $N$, a set of images $X$ with shape $(c, h, w)$, a trigger shape $(c_s, h_s, w_s)$ and a target label $t_s$. 
The goal is to check whether exists a trigger which successfully attacks every image in $X$. Algorithm $verifyX$ may have three outcomes. One is SAFE, i.e., there is no trigger such that backdoor attack succeeds on all the images in $X$. Another is UNSAFE, i.e., a trigger that can be used to successfully attack all images in $X$ is generated. The last one is UNKNOWN, i.e., we fail to establish either of the above results. 

In the following, we describe one concrete realization of the algorithm based on abstract interpretation, as shown in Algorithm~\ref{alg:verify1}.
At line 1, variable $\mathit{hasUnknown}$ is declared as a flag which is $true$ if and only if we cannot conclude whether there is a successful attack at a certain position. The loop from lines 2 to 15 tries every position for the trigger one by one. Intuitively, variable $\phi$ is the constraint that must be satisfied by a trigger to successfully attack every image in $X$. At line 3, we initialize $\phi$ to be $\phi_{pre}$, which is defined as follows: $\phi_{pre} \equiv \bigwedge_{j \in P(h_p, w_p)} lw_j \leq x_j \leq up_j$
where $j \in P(h_p, w_p)$ denotes that $j$ is an index (of an image pixel) in the trigger, 
$x_j$ is a variable denoting the value of the $j$-th pixel, $lw_j$ and $up_j$ are the (normalized) minimum (e.g., 0) and maximum (e.g., 1) value of feature $j$ in the image according to the input domain specified by the network $N$.
Intuitively, $\phi_{pre}$ requires that the pixels in the trigger must be within its domain. 

\begin{algorithm}[t]
\small
let $\mathit{hasUnknown} \leftarrow \mathit{false}$; \\
\ForEach{trigger position $(h_p, w_p)$} {
    let $\phi \leftarrow \phi_{pre}$;\\
    \ForEach{image I $\in$ X}{
        let $\phi_I \leftarrow attackCondition(N, I, \phi_{pre}, (c_s, h_s, w_s), (h_p, w_p), t_s)$; \\
        \If{$\phi_I$ is UNSAT}{
            $\phi \leftarrow \mathit{false}$; \\
            break;
        } \Else {
            $\phi \leftarrow \phi \land \phi_I$; \\
        }
    }
    \If{solving $\phi$ results in SAT or UNKNOWN}{
    \If {$opTrigger(N, X, \phi, (c_s, h_s, w_s), (h_p, w_p), t_s)$ returns a trigger} {
                \Return UNSAFE;
            }
    \Else{
            $\mathit{hasUnknown} \leftarrow \mathit{true}$;
            }
    }
}
\Return $\mathit{hasUnknown}$~?~UNKNOWN~:~SAFE;
\caption{$\mathit{verifyX(N, X, (c_s, h_s, w_s), t_s)}$}
\label{alg:verify1}
\end{algorithm}

Given a position, the loop from lines 4 to 10 constructs one constraint $\phi_I$ for each image $I$, which is the constraint that must be satisfied by the trigger to attack $I$. In particular, at line 5, function $attackCondition$ 
is called to construct the constraint. We present the details of this function in Section~\ref{abs-inter}. If $\phi_I$ is UNSAT (line 6), attacking image $I$ at position $(h_p, w_p)$ is impossible and we set $\phi$ to be $false$ and break the loop. Otherwise, we conjunct $\phi$ with $\phi_I$. 

After collecting one constraint from each image, we solve $\phi$ using a constraint solver. If it is not UNSAT (i.e., SAT or UNKNOWN), function $opTrigger$ is called to generate a trigger which is successful on all images in $X$ (if possible). Note that due to over-approximation, the model returned by the solver might be spurious. The details of function $opTrigger$ is presented in Section~\ref{sec:generation}. If a trigger is successfully generated, we return UNSAFE (at line 13, together with the trigger); otherwise, we set $\mathit{hasUnknown}$ to be $true$ and continue with the next trigger position. Note that we can return UNKNOWN at line 15 without missing any opportunity for verifying the backdoor absence. We instead continue with the next trigger location hoping a trigger may be generated successfully. After analyzing all trigger positions (and not finding a successful trigger), if $\mathit{hasUnknown}$ is $true$, we return UNKNOWN or otherwise SAFE.

\subsection{Abstract Interpretation} \label{abs-inter}
Function $attackCondition$ returns a constraint that must be satisfied such that the trigger with shape $(c_s, h_s, w_s)$ is successful on the image $I$ at position $(h_p, w_p)$. In this work, for efficiency reasons, it is built based on abstract interpretation techniques~\cite{singh2019abstract}. 
Multiple abstract domains have been proposed to analyze neural networks, such as interval~\cite{wang2018formal}, Zonotope~\cite{singh2018fast} and DeepPoly~\cite{singh2019abstract}. In this work, we adopt the DeepPoly abstract domain~\cite{singh2019abstract}, which is shown to balance between precision and efficiency. 

In the following, we assume each hidden layer in the network is expanded into two separable layers, one for the affine transformation and the other for the activation function transformation. We use $l$ to denote the number of layers in the expanded network,
$n_i$ to denote the number of neurons in layer $i$, and $x^I_{i,j}$ to denote the variable representing the $j$-th neuron in layer $i$ for the image $I$.
The constraint $\phi_I$ to be returned by function $attack(N, I, \phi_{pre}, (c_s, h_s, w_s), (h_p, w_p), t_s)$ is a conjunction of three parts.
$$\phi_I \equiv pre_I \wedge \mathcal{A_I} \wedge post_I$$
where $pre_I$ is the constraint on the input features according to the image $I$, i.e.,
$pre_I \equiv \phi_{pre} \wedge \left(\bigwedge_{j \in P(h_p, w_p)} x^I_{0,j} = x_j\right) \wedge \left(\bigwedge_{j \not \in P(h_p, w_p)} x^I_{0,j} = I[j]\right)$
where $j \not \in P(h_p, w_p)$ means that $j$ is not an index (of a pixel) of the trigger; $x^I_{0,j}$ is the variable that  represents the input feature $j$ (a.k.a.~neuron $j$ at the input layer) of the image $I$ and $I[j]$ is the (normalized) pixel value in the image at index $j$. Intuitively, the constraint $pre_I$ ``erases'' the pixels in the trigger, i.e., they can now take any value with their range, while the remaining pixels must have those value from the image. $post_I$ represents the condition for a successful attack. That is, the value of the target label (i.e., $x^I_{l-1,t_s}$) must be greater than the values of any other label, i.e.,
$post_I \equiv \bigwedge_{0 \leq j < n_{l-1} \wedge j \neq t_s} x^I_{l-1,t_s} > x^I_{l-1,j}$. 
    
More interestingly, $\mathcal{A_I}$ is a constraint that over-approximates the behavior of the neural network $N$ according to the DeepPoly abstract domain. That is, given the constraint on the input layer $pre_I$, a set of abstract transformers are applied to compute a linear over-approximation of every neuron in the next layer, every neuron in the layer after that, and so on until the output layer. The constraint computed on each neuron $x^I_{i,j}$ is of the form $ge^I_{i,j} \leq x^I_{i,j} \leq le^I_{i,j} \wedge lw^I_{i,j} \leq x^I_{i,j} \leq up^I_{i,j}$
where $ge^I_{i,j}$ and $le^I_{i,j}$ are two linear expressions constituted by variables representing neurons from the previous layer (i.e., layer $i-1$);  and $lw^I_{i,j}$ and $up^I_{i,j}$ are the concrete lower bound and upper bound of the neuron. Note that the abstract transformers are different for the activation function layer and affine layer. As the DeepPoly abstract transformers are not our contribution, we skip the details and refer the reader to~\cite{singh2019abstract} for details on the abstract transformers, including their soundness (i.e., they always over-approximate).

\begin{figure*}[t]
\centering
\includegraphics[scale=0.07]{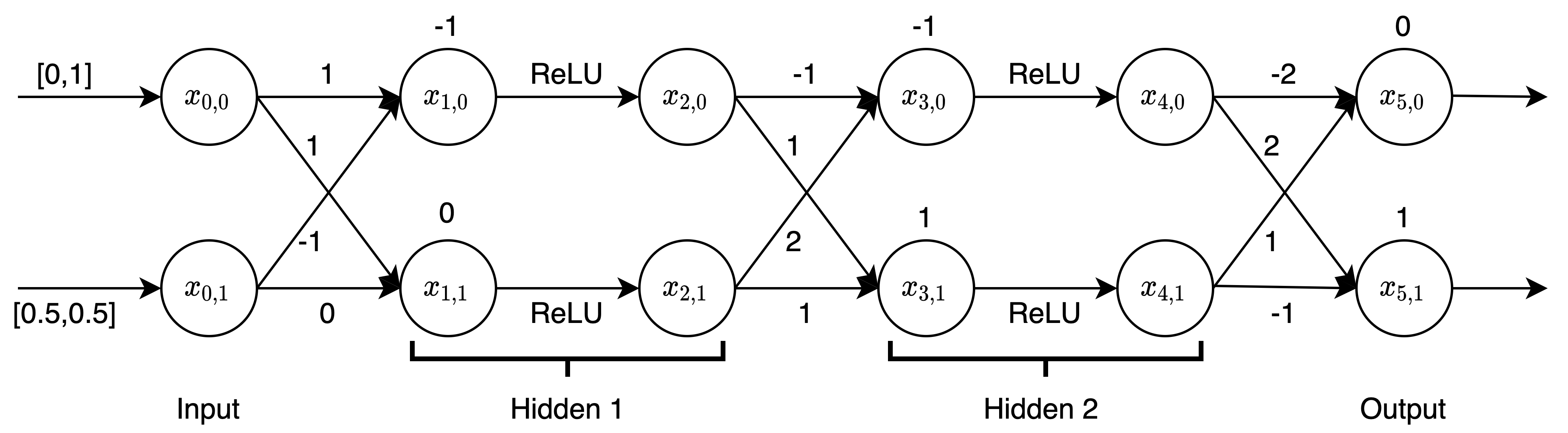}
\caption{An example of abstract interpretation}
\label{fig:mov_bd}
\end{figure*}

\begin{example}
Since it is too complicated to show the details of applying abstract interpretation to the neural network shown in Fig.~\ref{fig:network}, we instead construct a simple example as shown in Fig.~\ref{fig:mov_bd} to illustrate how it works. There are two features in this artificial image $I$, i.e., $x_{0,1}^I$ has a constant value of 0.5 and $x_{0,0}^I$ is the trigger whose value ranges from 0 to 1. That is,
$pre_I \equiv 0 \leq x_{0,0}^I \leq 1 \land x_{0,1}^I = 0.5$. 
After expanding the hidden layers, the network has 6 layers, each of which has 2 neurons.
Applying the DeepPoly abstract transformers from the input layer all the way to the output layer, we obtain the abstract states for the last layer.
Further, assume that the target label is 0. The constraint $post_I$ is thus as follows: $post_I \equiv x^I_{5,0} > x^I_{5,1}$. Solving the constraints returns SAT with $x^I_{0,0} = 0$. Indeed, with the stamped image $I_s = [0, 0.5]$, the output vector is $[1, 0]$. We thus identified a successful attack on the target label 0.
\end{example}

\noindent \emph{Optimization} Note that at line 6 of Algorithm~\ref{alg:verify1}, for each constraint $\phi_I$, we perform a quick check to see if the constraint is satisfiable or not. If $\phi_I$ is UNSAT, we can ignore the remaining images and analyze the next trigger position, which allows us to speed up the process. One naive approach is to call a solver on $\phi_I$, which would incur significant overhead since it could happen many times. To reduce the overhead, we propose a simple procedure to quickly check whether $\phi_I$ is UNSAT based solely on its abstract states at the output layer. That is, we check the satisfiability of the following constraint instead: 
$\bigwedge_{0 \leq j < n_{l-1} \wedge j \neq t_s} up^I_{l-1,t_s} > lw^I_{l-1,j}$. 
Recall that $up^I_{l-1,t_s}$ is the concrete upper bound of the neuron $t_s$ and $lw^I_{l-1,j}$ is the concrete lower bound of the  neuron $j$ at the output layer. Thus, intuitively, we check whether the concrete upper bound of the target label $t_s$ is larger than the concrete lower bound of every other label. If it is UNSAT, it is impossible to have the target label as the result and thus the attack would fail on the image $I$.
We then only call the solver on $\phi_I$ if the above procedure does not return UNSAT. Furthermore, the loop in Algorithm~\ref{alg:verify1} can be parallelized straightforwardly, i.e., by using a separate process to verify against a different trigger position. Whenever a trigger is found by any of the processes, the whole algorithm is then interrupted.

\subsection{Generating Backdoor Triggers} \label{sec:generation}
In the following, we present the details of function $opTrigger$, which intuitively aims to generate a trigger $S$ with shape $(c_s, h_s, w_s)$ at position $(h_p, w_p)$ for attacking every image $I$ in $X$ successfully. If the solver applied to solve $\phi$ at line 11 of Algorithm~\ref{alg:verify1} returns a model that satisfies $\phi$, we first check whether the model is indeed a trigger that successfully attacks every image in $X$. Due to over-approximation of abstract interpretation, the model might be a spurious trigger. If it is a real trigger, we return the model. Otherwise, we employ an optimization-based approach to generate a trigger.  

Given a network $N$, one image $I$, a target label $t_s$, and a position $(h_p,w_p)$, let $I_s$ is the stamped image generated from $I$ by stamping $I$ with the trigger at the position $(h_p,w_p)$. We generate a backdoor trigger $S$ by minimizing the following loss function.
\begin{align*}
loss(N, I, S, (h_p, w_p), t_s) = \left\{
\begin{array}{ll}
0 & \mbox{if $n_s > n_o$} \\
(n_o - n_s + \epsilon) & \mbox{otherwise}
\end{array}
\right.
\end{align*}
where $n_s = N(I_s)[t_s]$ is the output value of the target label; $n_o = {\max_{j \neq t_s}}~N(I_s)[j]$ is the maximum value of any label other than the target label; and $\epsilon$ is a small constant (e.g., $10^{-9}$). Note that the trigger $S$ is the only variable in the loss function. Intuitively, the loss function returns $0$ if the attack on $I$ by the trigger is successful. Otherwise, it returns a quantitative measure on how far the attack is from being successful on attacking $I$. Given a set of images $X$, the loss function is defined as the sum of the loss for each image $I$ in $X$:
$loss(N, X, S, (h_p,w_p), t_s) = \sum_{I \in X}loss(N, I, S, (h_p,w_p), t_s)$. 
The following optimization problem is then solved to find an attack which successfully attacks all images in $X$:
${\arg\min}_{S}~loss(N, X, S, (h_p,w_p), t_s)$.

\subsection{Correctness and Complexity}
In the following, we establish the correctness of Algorithm~\ref{alg:overall}.

\begin{lemma}
Given a neural network $N$, a set of images $X$, a trigger shape $(c_s, h_s, w_s)$, and a target label $t_s$, Algorithm~\ref{alg:verify1} (1) returns SAFE only if there is no backdoor attack which is successful on all images in $X$ with the provided trigger shape and target label; and (2) returns UNSAFE only if there exists a backdoor attack which is successful on all images in $X$ with the provided trigger shape and target label.
\end{lemma}
\begin{proof}
By~\cite{singh2019abstract}, function $attackCondition$ always returns a constraint which is an over-approximation of the constraint that must be satisfied such that the trigger is successful on image $I$.
Furthermore, Algorithm~\ref{alg:verify1} returns SAFE only at line 16, i.e., only if constraints that must be satisfied to attack all images in $X$ at each certain position are UNSAT. Thus, (1) is established. (2) is trivially established since we only return UNSAFE when a trigger that is successful on every provided image is generated. \hfill \qed
\end{proof}

The following establishes the soundness of our approach.

\begin{theorem}
Given a neural network $N$, a success rate $\theta$, a target label $t_s$, a trigger shape $(c_s, h_s, w_s)$, a type I error $\alpha$, a type II error $\beta$, and a half-width of the indifference region $\delta$, Algorithm~\ref{alg:overall} returns SAFE only if there is sufficient evidence (subject to type I error $\alpha$ and type II error $\beta$) that there is no backdoor attack with a success rate at least $\theta$ with the provided trigger shape and target label at the specified significance level.
\end{theorem}

\begin{proof}
If there is a backdoor attack with a success rate no less than $\theta$, given a set of randomly $K$ selected images, the probability of having an attack is no less than $\theta^K$ (since there is at least one backdoor attack with a success rate no less than $\theta$ and maybe more). Thus, the probability of not having an attack is no more than $1-\theta^K$. By the correctness of the SPRT algorithm, Algorithm~\ref{alg:overall} returns SAFE only if there is sufficient evidence that $H_0$ is true, i.e., the probability of not having an attack on a set of $K$ randomly selected images is more than $1-\theta^K$, implying it is sufficient evidence that there is no backdoor attack with success rate no less than $\theta$. The theorem holds. \hfill \qed
\end{proof}

Furthermore, it is trivial to show that Algorithm~\ref{alg:overall} returns UNSAFE only if there exists a backdoor attack which has a success rate at least $\theta$ with the provided trigger shape and target label.

In the following, we briefly discuss the complexity of our approach. It is straightforward to see that Algorithm~\ref{alg:verify1} always terminates if a timeout is imposed on solving the constraints and the optimization problems. Since we can always set a tight time limit on solving the constraints and the optimization problems, the complexity of the algorithm is determined mainly by the complexity of function $attackCondition$, which in turn is determined by the complexity of abstract interpretation.
The complexity of applying abstract interpretation with the DeepPoly abstract domain is $\mathcal{O}(l^2 \times n_{max}^3)$ where $l$ is the number of layers, and $n_{max}$ is the maximum number of neurons in any of the layers. Let $K$ be the number of images in $X$. Note that the number of trigger positions is $\mathcal{O}(h \times w)$, i.e., the size of an image. The best case complexity of Algorithm~\ref{alg:verify1} is $\mathcal{O}(l^2 \times n_{max}^3 \times h \times w)$ and the worst case complexity is $\mathcal{O}(l^2 \times n_{max}^3 \times K \times h \times w)$. 
We remark that in practice, $l$ typically ranges from 1 to 20; $n_{max}$ is often advised to be no more than the input size (e.g., from dozens to thousands usually); 
$K$ ranges from a few to hundreds; and $h \times w$ depends on the image resolution (e.g., from hundreds to millions). Thus, in general, Algorithm~\ref{alg:verify1} could be time-consuming in practice and we anticipate further optimization in future work. 

The complexity of Algorithm~\ref{alg:overall} is the complexity of Algorithm~\ref{alg:verify1} times the complexity of the SPRT algorithm. The complexity of the SPRT algorithm is in general hard to quantify and we refer the readers to~\cite{agha2018survey} for a detailed discussion.

\section{Implementation and Evaluation}
\label{eval}

We have implemented our approach as a self-contained analysis engine in the Socrates framework~\cite{pham2020socrates}. 
We use Gurobi~\cite{gurobi} to solve the constraints and use scipy~\cite{2020SciPy-NMeth} to solve the optimization problems. 

We collect a set of 51 neural networks. 45 of them are fully connected ones and are trained on the MNIST training set (i.e., a standard dataset which contains black and white images of digits). These networks have the number of hidden layers ranging from 3 to 5. For each network, the number of neurons in each of its hidden layers ranges from 10 to 50, i.e., 10, 20, 30, 40, or 50. To evaluate our approach on neural networks built with different activation functions, each activation function (i.e., ReLU, Sigmoid, and Tanh) is used in 15 of the neural networks.
Among the remaining six networks, three of them are bigger fully connected networks adopted from the benchmarks reported in~\cite{singh2019abstract}. They are all built with the ReLU activation function. For convenience, we name the networks in the form of \emph{f\_k\_n} where \emph{f} is the name of the activation function, \emph{k} is the number of hidden layers, and \emph{n} is the number of neurons in each hidden layer. 
The remaining three networks are convolutional networks (which are often used in face recognition systems) adopted from~\cite{singh2019abstract}. Although they have the same structure, i.e., each of them has two convolutional hidden layers and one fully connected hidden layer, they are trained differently. One is trained in the normal way; one is trained using DiffAI~\cite{mirman2018differentiable}, and the last one is trained using projected gradient descent~\cite{dong2018boosting}. These training methods are developed to improve the robustness of neural networks against adversarial attacks. Our aim is thus to evaluate whether they help to prevent backdoor attacks as well. We name these networks \emph{conv}, \emph{conv\_diffai}, and \emph{conv\_pgd}. 

We verify the networks against the backdoor trigger with shape $(1,3,3)$. 
All the networks are trained using clean data since we focus on verifying the absence of backdoor attacks.
They all have precision of at least 90\%, except \emph{Sigmoid\_4\_10} and \emph{Sigmoid\_5\_10}, which have precision of 81\% and 89\% respectively. 
In the following, we answer multiple research questions. All the experiments are conducted using a machine with 3.1Ghz 16-core CPU and 64GB RAM. All models and experiment details are at~\cite{backdoor}. \\

\noindent \emph{RQ1: Is our realization of $verifyX$ effective?} This question is meaningful as our approach relies on Algorithm $verifyX$. To answer this question, for each network, we select the first 100 images in the test set (i.e., a $K$ of 100 for Algorithm~\ref{alg:overall}, which is more than sufficient) and then apply Algorithm $verifyX$ with these images and each of the labels, i.e., 0 to 9. In total, we have 510 verification tasks. For each network, we run 10 processes in parallel, each of which verifies a separate target.
The only exception is the network \emph{ReLU\_3\_1024}, due to its complexity, we only run five parallel processes (since each process consumes a lot of resources).
In each verification process, we filter out those images which are classified wrongly by the network as well as the images which are already classified as the target label.

\begin{figure}[t]
\centering
\includegraphics[width=\linewidth]{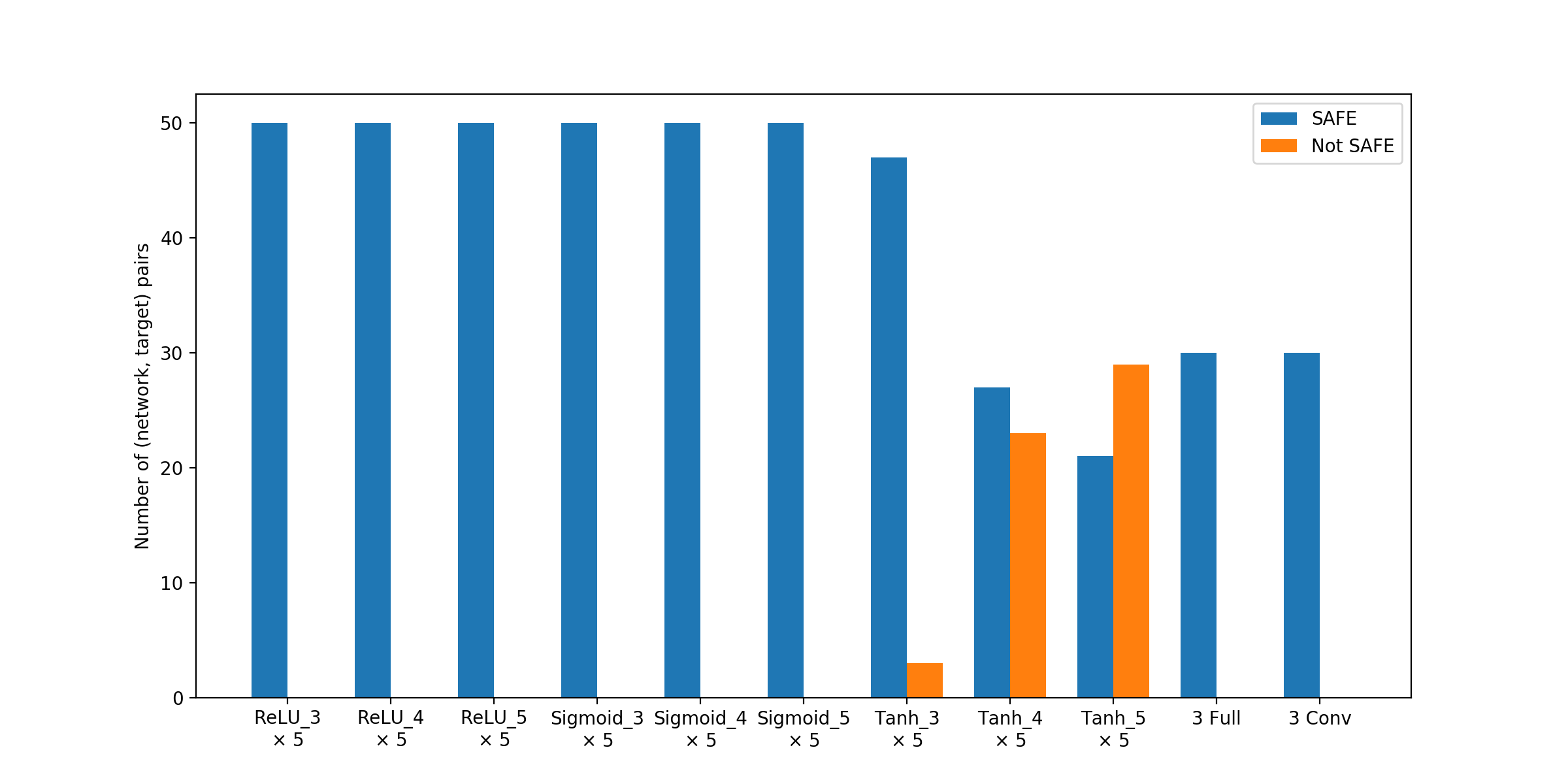}
\caption{Results of $verifyX$}
\label{fig:verifyI}
\end{figure}

Fig.~\ref{fig:verifyI} shows the results. The x-axis show the groups of the networks, e.g., \emph{ReLU\_3} means five fully connected networks using the ReLU activation function with three hidden layers; \emph{3 Full} and \emph{3 Conv} mean the three fully connected and the three convolutional networks adapted from~\cite{singh2019abstract} respectively. The y-axis shows the number of (network, target) pairs. Note that each group may contain a different number of pairs, i.e., the maximum values for the small network groups are 50, and the maximum values for the last two groups are 30.
First, we successfully verify 455 out of 510 verification tasks (i.e., 89\%) of them, i.e., the neural network is safe with respect to the selected images. It is encouraging to notice that the verified tasks include all models adopted from~\cite{singh2019abstract}, which are considerably larger (e.g., with 1024 neurons at each layer) and more complex (i.e., convolutional networks). Second, some networks are not proved to be safe with some target labels. It could be either there is indeed a backdoor trigger that we fail to identify (through optimization), or we fail to verify due to the over-approximation introduced by abstract interpretation. 
Lastly, with the same structure (i.e., the same number of hidden layers and the same number of neurons in each hidden layer), the networks using the ReLU and Sigmoid activation functions are more often verified to be safe than those using the Tanh activation function. This is most likely due to the difference in the precision of the abstract transformers for these functions.  
\\

\noindent \emph{RQ2: Can we verify the absence of backdoor attacks with a certain level of success rate?} To answer this question, we evaluate our approach on six networks used in RQ1, i.e., \emph{ReLU\_3\_10}, \emph{ReLU\_5\_50}, \emph{Sigmoid\_3\_10}, \emph{Sigmoid\_5\_50}, \emph{Tanh\_3\_10}, and \emph{Tanh\_5\_50}. These networks are chosen to cover a wide range of the number of hidden layers and the number of neurons in each layer, as well as different activation functions. Note that due to the high complexity of Algorithm~\ref{alg:overall} (which potentially applies Algorithm~\ref{alg:verify1} hundreds of times), running Algorithm~\ref{alg:overall} on all the networks evaluated in RQ1 requires an overwhelming amount of resources. \emph{Furthermore, since there is no existing work on backdoor verification, we do not have any baseline to compare with.}

Recall that Algorithm~\ref{alg:overall} has two important parameters $K$ and $\theta$, both of which potentially have a significant impact on the verification result. We thus run each network with four different settings, in which the number of images $K$ is set to be either 5 or 10, and the success rate $\theta$ is either 0.8 or 0.9. In total, with 10 target labels, we have a total of 240 verification tasks for this experiment. Note that some preliminary experiments are conducted before we select these two $K$ values. 

We use all the 10000 images in the test set as the image population and randomly choose $K$ images in each round of test. 
When a trigger is generated, the success rate of the trigger is validated on the images in the test set (after the above-mentioned filtering). Like in RQ1, we run each network with 10 parallel processes, each of which verifies a separate target. As the SPRT algorithm may take a very long time to terminate, we set a timeout for each verification task, i.e., 2 hours for those networks with three hidden layers, and 10 hours for those networks with five hidden layers. 
 
\begin{figure}[t]
\centering
\includegraphics[width=\linewidth]{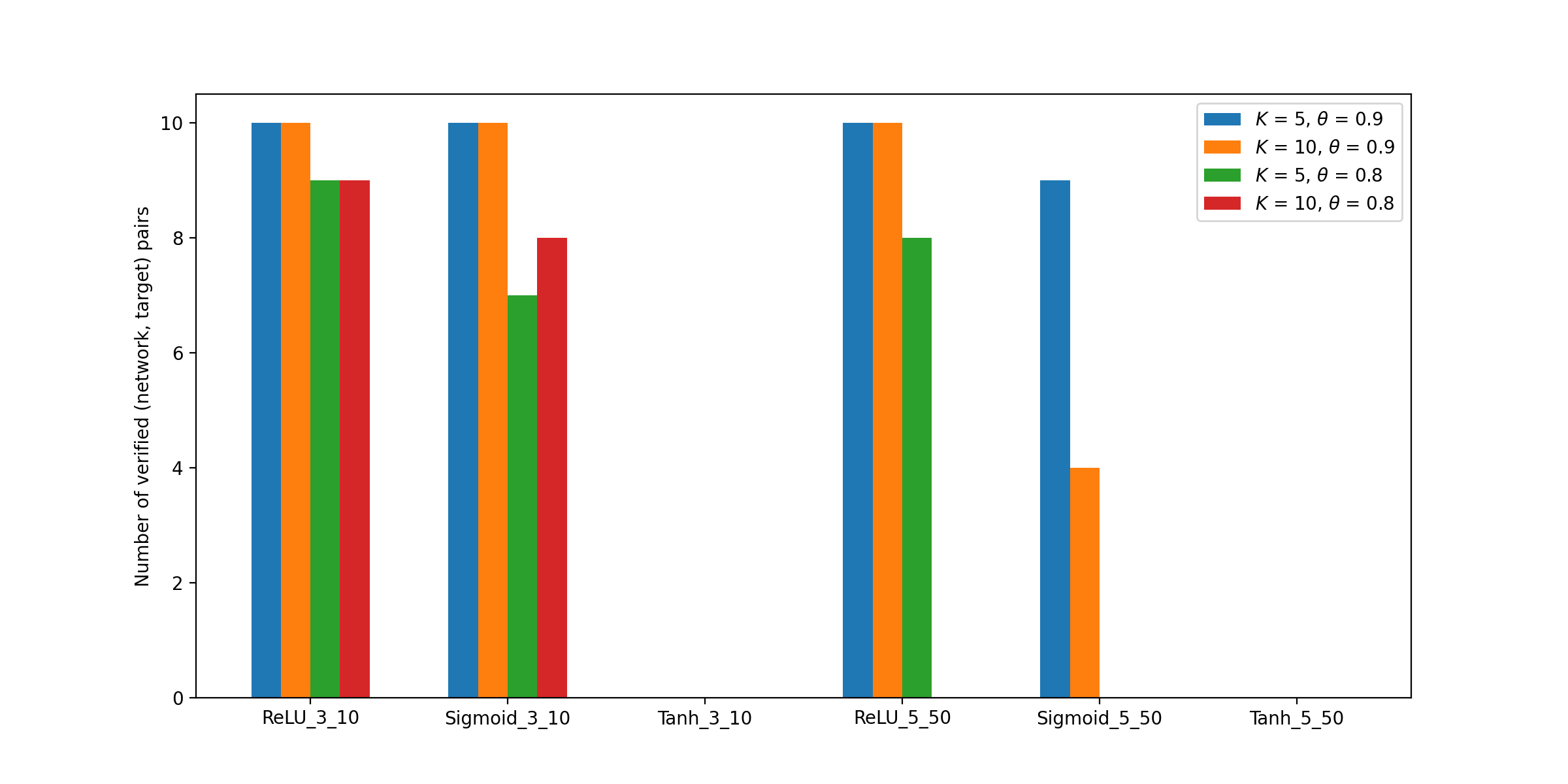}
\caption{Verification results}
\label{fig:verifyPr}
\end{figure}

The results are shown in Fig.~\ref{fig:verifyPr}. The x-axis shows the networks, the y-axis shows the number of verified pairs of network and target label.
We have multiple observations based on the experiment results. First, a quick glance shows that with the same structure and hypothesis testing parameters, more networks built with the ReLU activation function are verified than those built with the Sigmoid and Tanh functions. 
Second, we notice that the best result is achieved with $K = 5$ and $\theta = 0.9$. With these parameter values, we can verify that three networks \emph{ReLU\_3\_10}, \emph{ReLU\_5\_50}, and \emph{Sigmoid\_3\_10} are safe with respect to all the target labels and the network \emph{Sigmoid\_5\_50} is safe with respect to nine over 10 target labels. If we keep the same success rate as 0.9 and increase the number of images $K$ from 5 to 10, we can see that the number of verified cases in the network \emph{Sigmoid\_5\_50} decreases. This is because when we increase the number of images that must be attacked successfully together, the probability that we do not have the attack increases, which means we need more rounds of test to confirm the hypothesis $H_0$ and so the verification process for the network \emph{Sigmoid\_5\_50} times out before reaching the conclusion. We have a similar observation when we keep the number of images $K$ at 5 but decrease the success rate from 0.9 to 0.8. When the success rate decreases, the probability of not having the attack increases, which requires more tests to confirm the hypothesis $H_0$. As a result, for all these four networks, there are multiple verification tasks that time out before reaching the conclusion. However, we notice that there is an exception when we keep the success rate as 0.8 and increase the number of images from 5 to 10. While the number of verified cases for the network \emph{ReLU\_5\_50} decreases (which can be explained in the same way as above), the number of verified cases for the network \emph{Sigmoid\_3\_10} increases (and the results for the other two networks do not change). Our explanation is that when we increase the number of images $K$ to 10, it is easier for the Algorithm~\ref{alg:verify1} to conclude that there is no attack, and so the Algorithm~\ref{alg:overall} still collects enough evidence to conclude $H_0$. On the other hand, with the number of images is 5,  Algorithm~\ref{alg:verify1} may return a lot of UNKNOWN (due to spurious triggers), and so the hypothesis testing in the Algorithm~\ref{alg:overall} goes back and forth between the two hypotheses $H_0$ and $H_1$ and eventually times out. 

A slightly surprising result is obtained for the network \emph{Tanh\_3\_10}, i.e., our trigger generation process generates two triggers for the target labels 2 and 5 when the success rate is set to be $0.8$.
This is surprising as these networks are not generated with backdoor attack. This result can be potentially explained by the combination of the relatively low success rate (i.e., 0.8) and the phenomenon known as universal adversarial perturbations~\cite{moosavi2017universal}. With the returned triggers, the users may want to investigate the network further and potentially improve it with techniques such as robust training~\cite{mirman2018differentiable,dong2018boosting}. \\

\begin{figure}[t]
\centering
\includegraphics[scale=0.5]{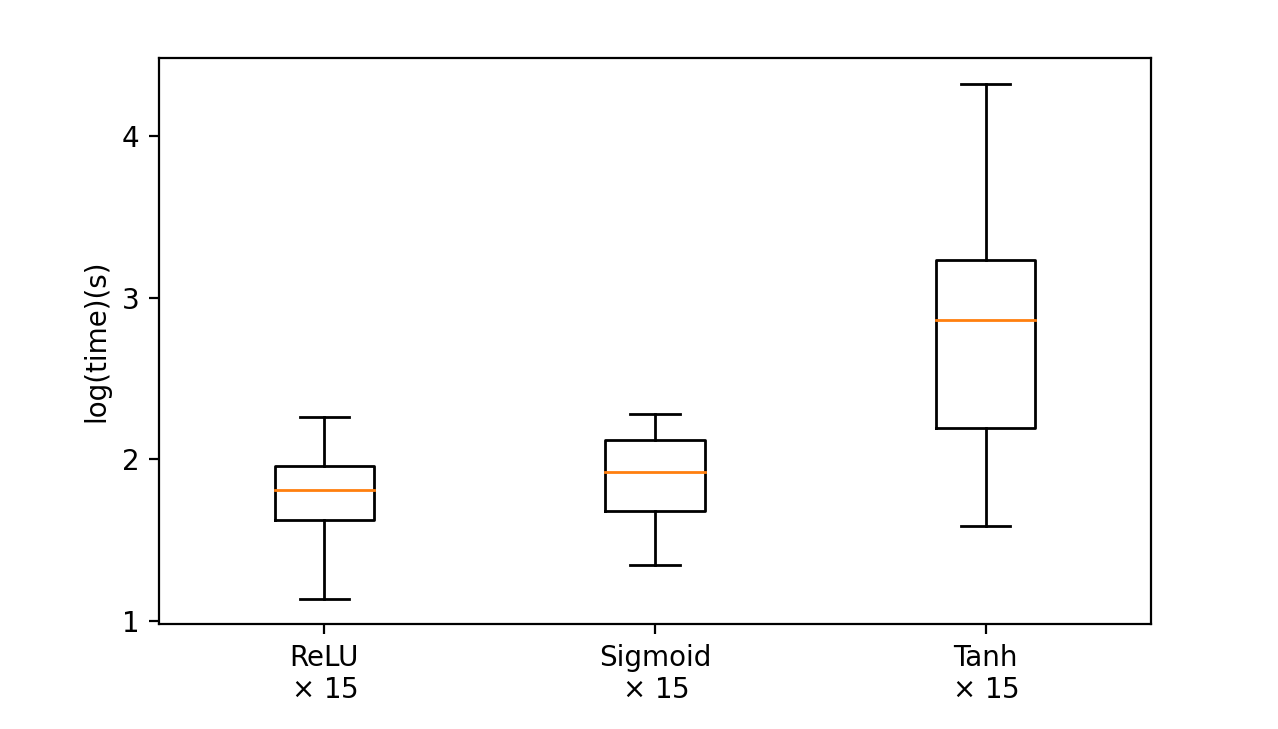}
\caption{The running time of the experiments in RQ1 with benchmark networks}
\label{fig:time}
\end{figure}
\noindent \emph{RQ3: Is our approach efficient time-wise?}
To answer this question, we collect the wall-clock time to run the experiments in RQ1 and RQ2. For each network, we record the average running time for 10 different target labels. The results for 45 small networks are shown in Fig.~\ref{fig:time}. The x-axis shows the groups of 15 networks categorized based on their activation functions and the y-axis shows the logarithmic scale of the running time in the form of boxplots (where the box shows the result of 25 percentile to 75 percentile, the bottom and top lines are the minimum and maximum, and the orange line is median).
The execution time ranges from 14 seconds to less than 6 hours for these networks. Furthermore, we can see that there is not much difference between the running time of the networks using the ReLU and Sigmoid activation functions. However, the running time of the networks using the Tanh function is one order of magnitude larger than those of the ReLU and Sigmoid networks. The reason is that the Tanh networks have many non-safe cases (as shown in Fig.~\ref{fig:verifyI}) and, as a result, 
the verification process needs to check more images at more trigger positions. The running time of those networks adopted from~\cite{singh2019abstract} ranges from more than 5 minutes to less than 4 hours, as shown in Table~\ref{tab:timeDeepPoly}.
Finally, the running time for each network in RQ2 (i.e., the time required to verify the networks against backdoor attacks) according to different settings is shown in Table~\ref{tab:timePr}.


\begin{table}[t]
    \centering
    \setlength{\tabcolsep}{18pt}
    \begin{tabular}{llll}
        \toprule
       \textbf{Network}  &  \textbf{Time}  & \textbf{Network} & \textbf{Time} \\
       \midrule
       ReLU\_3\_1024  &  237m 24s & conv         & 194m 30s \\
       ReLU\_5\_100   &  5m 38s   & conv\_diffai & 111m 12s \\
       ReLU\_8\_200   &  48m 34s  & conv\_pgd    & 190m 19s \\
       \bottomrule
    \end{tabular}
    \caption{The running time of the experiments in RQ1 with networks adapted from~\cite{singh2019abstract}}
    \label{tab:timeDeepPoly}
\end{table}

\begin{table}[t]
    \centering
    \setlength{\tabcolsep}{15pt}
    \begin{tabular}{lllll}
        \toprule
       \textbf{Network}  &  $K = 5$  & $K = 10$ & $K = 5$ & $K = 10$ \\
        &  $\theta=0.9$  & $\theta=0.9$ & $\theta=0.8$ & $\theta=0.8$ \\
       \midrule
       ReLU\_3\_10     &  31m 31s  & 46m 39s  & 55m 44s  & 68m 54s \\
       ReLU\_5\_50     &  341m 36s & 493m 30s & 551m 40s & 600m 0s \\
       Sigmoid\_3\_10  &  46m 43s  & 59m 28s  & 92m 34s  & 93m 21s \\
       Sigmoid\_5\_50  &  476m 38s & 588m 25s & 600m 0s  & 600m 0s \\
       Tanh\_3\_10     &  114m 2s  & 105m 18s & 50m 58s  & 26m 4s \\
       Tanh\_5\_50     &  600m 0s  & 600m 0s  & 600m 0s  & 600m 0s \\
       \bottomrule
    \end{tabular}
    \caption{The running time of the experiments in RQ2}
    \label{tab:timePr}
\end{table}


\noindent \emph{RQ4: Can our approach generate backdoor triggers?}
Being able to generate counterexamples is a part of a useful verification method. We conduct another experiment to evaluate the effectiveness of our backdoor trigger generation approach. We train a new set of 45 networks that have the same structure as those used for answering RQ1. The difference is that this time each network is trained to contain backdoor through data poisoning. In particular, for each network, we randomly extract 20\% of the training data, stamp a white square with shape $(1,3,3)$ in one corner of the images, assign a random target label, and then train the neural network from scratch with the poisoned training data. While such an attack is shown to be effective~\cite{gu2017badnets}, it is not guaranteed to be always successful on a randomly selected set of images. Thus, we do the following to make sure that there exists a trigger for a set of selected images. From 10000 images in the test set, we first filter out those images which are classified wrongly or already classified with the target label. The remaining images are collected into a set $X_0$. Next, to make sure that the selected images have a high chance of being attacked successfully, we apply another filter on $X_0$. This time, we stamp each image in $X_0$ with a white square at the same trigger position as we poison the training data. We then keep the image if its stamped version is classified by the network with the target label. The remaining images after the second filter are collected into another set $X$. We apply our approach, in particular, the backdoor trigger generation on $X$, if $|X| \div |X_0| \geq 0.8$, i.e., the backdoor attack has a success rate of 80\%.

\begin{figure}[t]
\centering
\includegraphics[width=\linewidth]{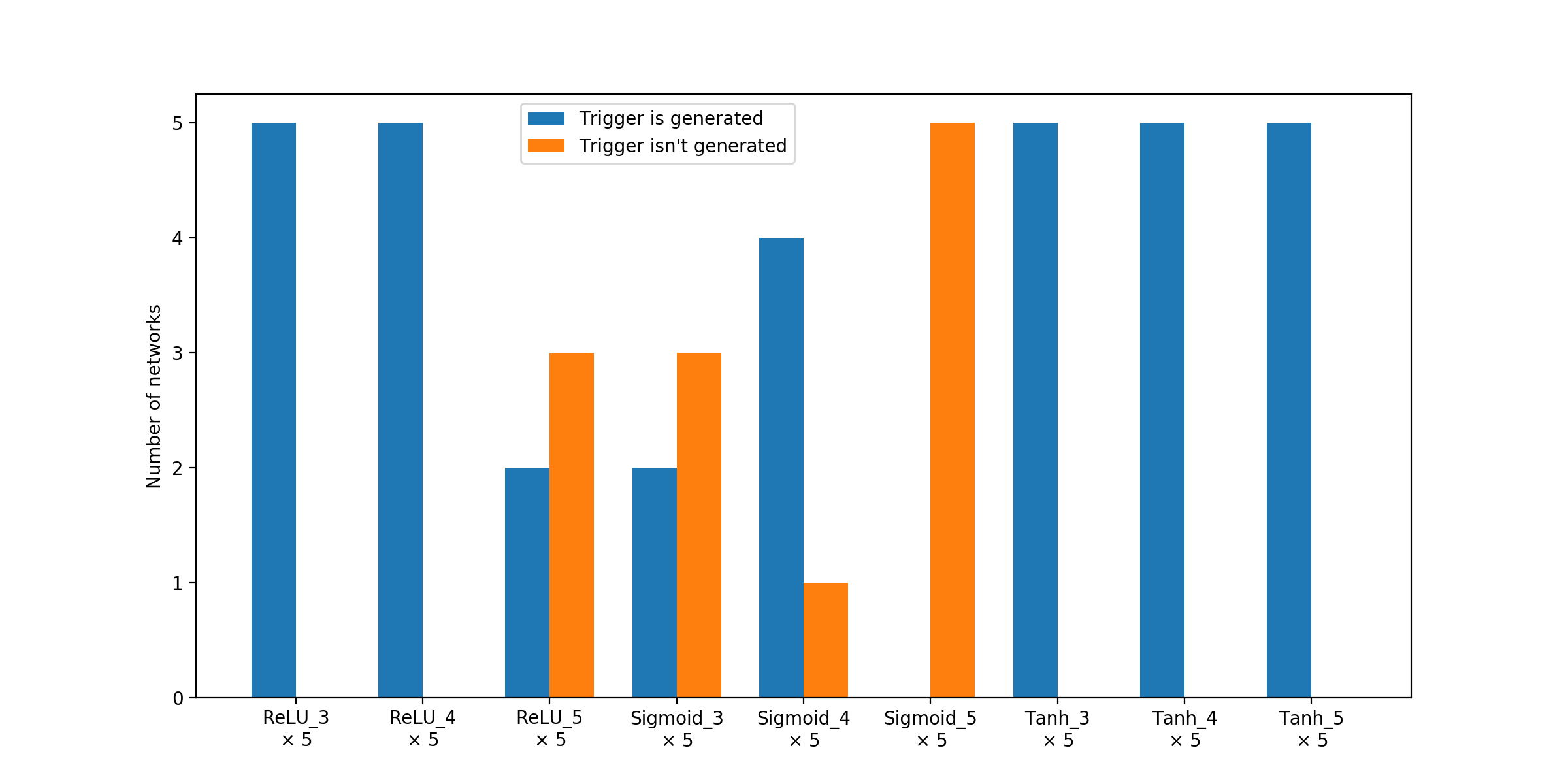}
\caption{The results of backdoor trigger generation}
\label{fig:atk}
\end{figure}

The results are shown in Fig.~\ref{fig:atk} in which the y-axis shows the number of networks. The timeout is set to be 120 seconds. Among the 45 networks, we can see that a trigger is successfully generated for 33 (i.e., 73\%) of the networks. A close investigation of these networks shows that the generated trigger is the exact white square that is used to stamp the training data. There are 12 networks for which the trigger is not generated. We investigate these networks and see that they are either too biased (i.e., classifying every image with the target label and thus $|X_0| = 0$) or the attack on these networks does not perform well (i.e., $|X| \div |X_0| < 0.8$). In other words, the backdoor attack on these networks failed and, as a result, the generation process does not even begin with these networks. In a nutshell, we successfully generate the trigger for every successful backdoor attack. Finally, note that the running time of the backdoor generation process is reasonable (i.e., on average, 50 seconds to generate a backdoor trigger for one network) and thus it does not affect the overall performance of our verification algorithm.


\section{Related Work}
\label{related}

The work which is closest to ours is~\cite{wang2020certifying} in which Wang \emph{et al.} aim to certify neural networks' robustness against backdoor attack using randomized smoothing. However, there are many noticeable differences between their approach and ours. First, while our work focuses on verifying the absence of backdoor, their work aims to certify the robustness of individual images based on the provided training data and learning algorithm (which can be used to implicitly derive the network). Second, by using random noises to estimate the networks' behaviors, their approach can only obtain very loose results. As shown in their experiments, they can only certify the robustness against backdoor attack with triggers contains two pixels and on a ``toy'' network with only two layers and two labels, after simplifying the input features by rounding them into $0$ or $1$. Compare to their approach, our approach can apply to networks used to solve real image classification problems as shown in our experiments. In fact, the following is stated in their abstract: \emph{``However, we also find that existing randomized smoothing methods have limited effectiveness at defending against backdoor attacks, which highlight the needs of new theory and methods to certify robustness against backdoor attacks''}.

Our work is closely related to a line of work on verifying neural networks. Existing approaches mostly focus on local robustness property and can be roughly classified into two categories: exact methods and approximation methods.
The exact methods aim to model the networks precisely and solve the verification problem using techniques such as mixed-integer linear programming~\cite{tjeng2017evaluating} or SMT solving~\cite{katz2017reluplex,ehlers2017formal}. On the one hand, these approaches can guarantee sound and complete results in verifying neural networks. On the other hand, they often have limited scalability and thus are limited to small neural networks. Moreover, these approaches have difficulty in handling activation functions except the ReLU function.

In comparison, the approximation approaches over-approximate neural network behavior to gain better scalability. AI$^2$~\cite{gehr2018ai2} is the first work pursuing this direction using the classic abstract interpretation technique. After that, more researchers try to explore different abstract domains for better precision without sacrificing too much scalability~\cite{singh2018fast,singh2019abstract,singh2019beyond}.
Besides that, some authors adopt linear approximation~\cite{weng2018towards}, duality~\cite{wong2017provable,dvijotham2018dual}, Lipschitz continuity~\cite{weng2018towards,ruan2018reachability}, semidefinite relaxations~\cite{raghunathan2018semidefinite}, and  star-based approach~\cite{tran2019star,tran2020verification} in analyzing neural networks.
In general, the approximation approaches are more scalable than the exact methods, and they are capable of handling activation functions such as Sigmoid and Tanh. However, due to the over-approximation, these methods may fail to verify a valid property.

We also notice that it is possible to incorporate abstraction refinement to the approximation methods and gain better precision, for instance, by splitting an abstraction into multiple parts to reduce the imprecision due to over-approximation. There are many works~\cite{wang2018formal,wang2018efficient,lu2019neural} which fall into this category. We remark that our approach is orthogonal to the development of sophisticated verification techniques for neural networks. 

Finally, our approach, especially the part on backdoor trigger generation, is related to many approaches on generating adversarial samples for neural networks. Some representative approaches in this category are FGSM~\cite{goodfellow2014explaining}, JSMA~\cite{papernot2016limitations}, and C\&W~\cite{carlini2017towards} which aim to generate adversarial samples to violate the local robustness property, and~\cite{zhang2020white} which aims to violate fairness property.
\section{Conclusion}
\label{conclusion}

In this work, we propose the first approach to formally verify that a neural network is safe from backdoor attacks. We address the problem on how to verify the absence of a backdoor that reaches a certain level of success rate. Our approach is based on abstract interpretation and we provide an implementation based on DeepPoly abstract domain. The experiment results show the potential of our approach. In the future, we intend to extend our approach with more abstract domains as well as improve the performance to verify more real-life networks. Besides that, we also intend to apply our approach to verify the networks designed for other tasks, such as sound or text classification.

\subsubsection*{Acknowledgements}
This research is supported by the Ministry of Education, Singapore under its Academic Research Fund Tier 3 (Award ID: MOET32020-0004). Any opinions, findings and conclusions or recommendations expressed in this material are those of the author(s) and do not reflect the views of the Ministry of Education, Singapore. This research is also partly supported by the Starry Night Science  Fund of Zhejiang University Shanghai Institute for Advanced Study, Grant No. SN-ZJU-SIAS-001.

\bibliographystyle{IEEEtran}
\bibliography{IEEEabrv,ref}

\begin{subappendices}
\renewcommand{\thesection}{\Alph{section}}%
\section{Discussion} \label{discussion}
Our approaches are designed to verify the absence of input-agnostic (i.e., not input-specific) backdoor attacks as presented in Section~\ref{overview}. In the following, we briefly review other backdoor attacks and discuss how to extend our approach to support them.

In~\cite{gu2017badnets}, Gu \emph{et al.} described a backdoor attack which, instead of forcing the network to classify any stamped image with the target label, only alters the label if the original image has a specific ground truth label $t_i$ (e.g., Bob with the trigger will activate the backdoor and be classified as Alice the manager). Our verification approach can be easily adapted to verify the absence of this attack by focusing on images with label $t_i$ in Algorithm~\ref{alg:overall} and Algorithm~\ref{alg:verify1}. 

Another attack proposed in~\cite{gu2017badnets} works by reducing the performance (e.g., accuracy) of the neural network on the images with a specific ground truth label $t_i$, i.e., given an image with ground truth label $t_i$, the network will classify the stamped image with some label $t_s \neq t_i$. The attack can be similarly handled by focusing on images with ground truth label $t_i$, although due to the disjunction introduced by $t_s \neq t_i$, the constraints are likely to be harder to solve. 
That is, we can focus on images with ground truth label $t_i$, and define an attack to be successful if $L(N(I_s)) \neq t_i$ is satisfied.

In~\cite{liu2017trojaning}, Liu \emph{et al.} proposed to use backdoor triggers with different shapes (i.e., not just in the form of a square or a rectangle). If the user is aware of the shape of the backdoor trigger, a different trigger can be used as input for Algorithm~\ref{alg:overall} and Algorithm~\ref{alg:verify1} and the algorithms would work to verify the absence of such backdoor. Alternatively, the users can choose a square-shaped backdoor trigger that is larger enough to cover the actual backdoor trigger, in which case our algorithms would remain to be sound, although it might be inconclusive if the trigger is too big. 

Multiple groups~\cite{shafahi2018poison,turner2018clean,barni2019new,liu2020reflection} proposed the idea of poisoning only those samples in the training data which have the same ground truth label as the target label to improve the stealthiness of the backdoor attack. This type of attack is designed to trick the human inspection on the training data, and so does not affect our verification algorithms.

In this work, we consider a specific type of stamping, i.e., the backdoor trigger replaces the part of the original clean image. Multiple groups~\cite{chen2017targeted,liu2017trojaning} proposed the use of the blending operation as a way of `stamping', i.e., the features of the backdoor trigger are blended with the features of the original images with some coefficients $\alpha$. This is a form of input-specific backdoor, the trigger is different for different images. To handle such kind of backdoor attacks, one way is to modify the constraint $pre_I$ according to the blending operation (assuming that $\alpha$ is known).
Since the blending operation proposed in~\cite{chen2017targeted,liu2017trojaning} is linear, we expect this would not introduce additional complexity to our algorithms. 

Input-specific triggers, in general, may pose a threat to our approach. First, some input-specific triggers~\cite{liu2017trojaning,liu2020reflection} cover the whole image, which is likely to make our approach inclusive due to false alarms resulted from over-approximation. Second, it may not be easy to model some of the input-specific triggers in our framework. For instance, Liu \emph{et al.}~\cite{liu2020reflection} recently proposed to use reflection to create stamped images that look natural. Modeling the `stamping' operation for this kind of attack would require us to know where the reflection is in the image, which is highly non-trivial. However, it should also be noted that input-specific triggers are often not as effective as input-agnostic triggers, e.g., the reflection-based attack reported in~\cite{liu2020reflection} are hard to reproduce. Furthermore, as discussed in Section~\ref{overview}, backdoor attack with input-specific triggers is an attacking method that is more powerful than adversarial attacks, and the problem of verifying the absence of backdoor attack with input-specific triggers is not yet clearly defined. 
\end{subappendices}

\end{document}